## ARTICLE

# Spatially-resolved insulator-metal transition for rewritable optical gratings

Yuliang Chen[1], Changlong Hu[1], Liyan Xie[2,4], Xiaoyu Zhou[1], Bowen Li[1], Hui Ren[1], Liang Li[1], Guobin Zhang[1,3], Jun Jiang[2✉] & Chongwen Zou [1✉]

Optical relief gratings are usually composed of physical grooves with a constant periodicity, and typically suffer from light scattering, are mechanically fragile and are single function. Here, we develop $WO_3$-based gratings by using a recently reported electron-proton synergistic doping route under ambient conditions. This doping strategy is compatible with conventional ultraviolet photolithography, and we show that it induces a selective insulator-metal phase transition and coloration in $WO_3$, with spatial-resolution up to micron-scale. Due to the electrochromic-induced-contrast, a $WO_3$ volume phase grating without grooves and a $WO_3$ relief grating with tunable periodicity are demonstrated. Both gratings can be rewritten after a reset procedure by annealing in air. Our experiments demonstrate $WO_3$–based gratings and an attractive technique for rewritable oxides.

[1] National Synchrotron Radiation Laboratory, University of Science and Technology of China, Hefei, China. [2] Hefei National Laboratory for Physical Sciences at the Microscale, Collaborative Innovation Center of Chemistry for Energy Materials, CAS Center for Excellence in Nanoscience, School of Chemistry and Materials Science, University of Science and Technology of China, Hefei, China. [3] Anhui Laboratory of Advanced Photon Science and Technology, University of Science and Technology of China, Hefei, China. [4] Present address: Key Laboratory of the Ministry of Education for Advanced Catalysis Materials, College of Chemistry and Life Sciences, Zhejiang Normal University, Jinhua, China. ✉email: jiangj1@ustc.edu.cn; czou@ustc.edu.cn

Grating has a periodic structure that splits and diffracts light into several beams traveling in different directions. As one of the most important components in optics, gratings are commonly used in monochromators and spectrometers[1]. The classical relief gratings can be divided into two basic categories: ruled and holographic gratings, which always consist of physical grooves[2]. Though the fabrications of traditional gratings are mature, it is still facing great challenges. For example, the random errors and irregularities of grooves inevitably introduce stray light. Except simple lines, it is difficult to include complicated patterns into a single grating. Moreover, the traditional relief gratings usually have dense lines (>1000 per millimeter), which means they are fragile, even fingerprints or the slightest contact with any abrasive material would damage them. In addition, once a relief grating is fabricated, it has a constant period and other optical properties, which makes it difficult to undergo further modulation or reprocessing.

On the other hand, the common materials used for the production of typical relief gratings, such as glass, silicon, copper, or photoresist, do not have abundant properties to be modulated, which limits multifunction of grating devices[2–6]. Currently, metal oxides have attracted tremendous interest due to their controllable properties, such as metal–insulator transition (MIT) and electrochromism, which can be effectively triggered by hydrogenation-induced electron doping[7–9]. Tungsten trioxide ($WO_3$) crystallizes in the distorted $ReO_3$ type of structure, a perovskite-type crystal ($ABO_3$) with vacant A sites[10]. It is relatively easy to intercalate protons to occupy the A sites forming tungsten bronzes[11,12]. The protons doped $WO_3$ accompanying pronounced MIT (more than six orders of magnitude) and coloration (from transparent to dark blue) show broad potential in smart windows, artificial synaptic devices[13–16].

In this study, we developed rewritable $WO_3$-based gratings by using the recently proposed doping method, i.e., electron–proton synergistic doping route[17]. This doping technique can be conducted under ambient conditions, which proved to be compatible with conventional ultraviolet photolithography here. Thus, the spatially-resolved MIT and coloration in hydrogenated $WO_3$ were feasible up to micrometer scale. Based on this unique doping technique, we achieved a planar grating without physical grooves, which can be taken as a kind of new volume phase grating with intrinsically low scattering[18,19]. In addition, a $WO_3$ relief grating with tunable period was also fabricated by the selective synergistic doping. More importantly, these $WO_3$-based gratings can be reprocessed after a "reset" process by annealing in air, making the devices controllable and reusable. It is believed that the proposed $WO_3$-based gratings here will be promising alternatives of typical gratings in the future. This advanced doping technology opens up alternative approaches for developing not only optical devices, but also rewritable ions devices and integrated circuits for various oxide electronics.

## Results and discussion

**Electron–proton synergistic doping in $WO_3$ system.** Figure 1a shows a schematic diagram of electron–proton synergistic doping route. When an oxide contacted with a proper metal particle are immersed into an acid solution, the electrons supplied by metal (yellow particle) and the free protons in acid solution can be doped into oxide together rather than the oxide being corroded by acid. This electron–proton synergistic effect was firstly reported in our previous work[17], which was due to the variation of Fermi level ($E_F$) between the oxide and metal as shown in Fig. 1b. When the $E_F$ level of metal was higher than that of oxide, the electrons in metal would flow into oxide for balancing their $E_F$ levels. The interfacial charge transfer will not be maintained due to the electrostatic screen effect. While if this metal/oxide contact was put into an acid solution, the free protons in the solution would be attracted into oxide, forming the H-doped oxide layer. In addition, the protons insertion into oxide would metallize the contact area and break the electrostatic screen effect, which drove continuous electron doping into the oxide. The more detailed discussion was available in Supplementary Fig. 1.

Based on above model, we predicted that $WO_3$ could be a good sample due to its high work function[20–22], 6.59 eV, as shown in the histogram of Fig. 1c. Indeed, when a $WO_3$ film was touched by a zinc particle (~2 mm size) with the work function of 4.33 eV and put into a diluted $H_2SO_4$ solution, a dark blue region was formed and spread quickly to the whole film in Fig. 1d. An animated process is in Supplementary Video 1. Because the spreading speed is limited, the doping region can be controlled and selected easily as shown in the scheme of Fig. 1e, a zinc pen was used for writing the characters of "$WO_3$" in a piece of $WO_3$ film (Fig. 1f), see more details in Supplementary Fig. 2 and Supplementary Video 2. The thick/thin lines in Fig. 1f should be attributed to slow/quick handwriting speed due to the prominent diffusion. More importantly, the coloration in $WO_3$ film was reversible. The "$WO_3$" characters were erased by 300 °C annealing in air as shown in Fig. 1g. The animated process is in Supplementary Video 3.

**Experimental and calculated results for $WO_3$ system.** The H atoms existed in the samples were examined by secondary-ion mass spectrum (SIMS) in Fig. 2a. The results showed that after the synergistic doping, the $WO_3$ film was hydrogenated into $H_xWO_3$ effectively, which had the highest H concentration (red curve) compared with the pristine $WO_3$ (hyacinth curve) and restored $WO_3$ (green curve)[15]. In contrast, O element was pretty constant for these three samples in Supplementary Fig. 3, thus the possible oxygen vacancies were excluded as same as the synergistic effect happened in $VO_2$ system[11,17,23]. To estimate the value of $x$ in $H_xWO_3$ film, the carrier density was $2.1 \times 10^{21}$ cm$^{-3}$ derived from the Hall measurement corresponding to $x = \sim 0.1$ (Supplementary Fig. 4)[11,12,24]. Any Zn element cannot be traced no matter at surface or in volume of $H_xWO_3$ (Supplementary Table 1). It was reported that for many H-doped insulators, their pristine characteristic peaks in Raman spectrum were usually changed even disappeared because of hydrogenation[25,26]. In Fig. 2b, it was observed that the 802 cm$^{-1}$ peak belonged to pristine $WO_3$ film was really disappeared in $H_xWO_3$ film[27]. Even the peaks of the sapphire substrate cannot be distinguished in Fig. 2b, which may be due to that the applied 633 nm excitation laser cannot go through the $H_xWO_3$ film. To verify this hint, we conducted UV–Vis–infrared transmission tests. Indeed, the great difference of the transmission for these three samples were observed in Fig. 2c. The pristine $WO_3$ film showed very high visible and near-infrared transmission, while the transmission of the hydrogenated $H_xWO_3$ film was near zero, except the low transmission in the range from 350 to 530 nm. This transmission property was quite consistent with the observation in Fig. 1d, which showed that the $H_xWO_3$ film was heavily dark blue by eyesight.

As we know, infrared is usually reflected by metal[28], which implies a MIT between pristine $WO_3$ and $H_xWO_3$. A more convinced electric characterization was conducted. Figure 2d showed the resistance–temperature ($R$–$T$) measurement for the pristine $WO_3$ film and the hydrogenated one. The resistance of pristine $WO_3$ film (~10$^8$ Ω) was decreased up to six orders of magnitude if hydrogenated to $H_xWO_3$ (~10$^2$ Ω), indicating a pronounced MIT. It also pointed out that the metallic $H_xWO_3$ was quite stable at ambient according to the $R$–$T$ cycle tests in the inset of Fig. 2d, which indicated the resistance of the $H_xWO_3$ film showed little changes even heated up to 100 °C conforming to the solid non-volatility of coloration at room temperature (Supplementary Fig. 5). Keeping the $H_xWO_3$ film in inert atmospheres or vacuum contributed to improving this non-volatility[16]. In addition, there was no noticeable variation in the scan electron microscope (SEM) results for the surface and cross-section scans for these film samples (Supplementary Fig. 6). From the X-ray diffraction (XRD) results, it was observed that the three films were all amorphous (Supplementary Fig. 7), since there were no related diffraction peaks for $WO_3$ in the curves, see more details in "Methods" section.

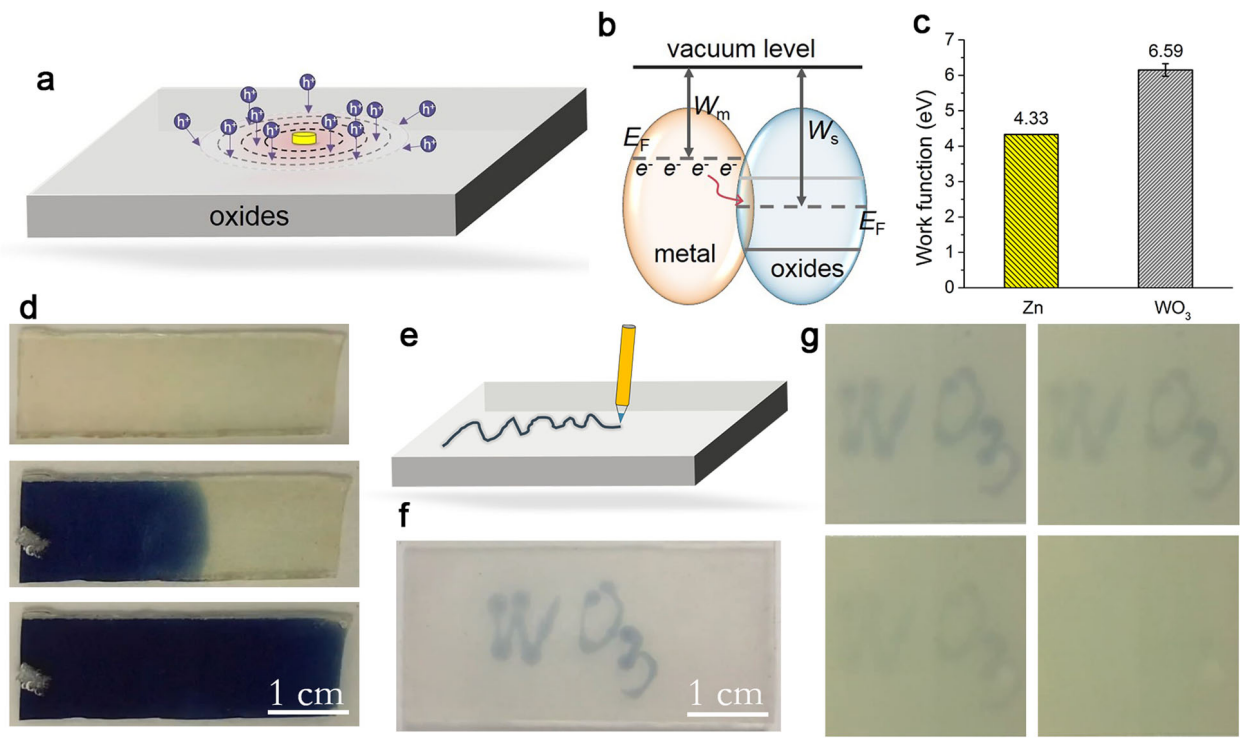

**Fig. 1 The electron–proton synergistic doping for WO₃ film. a** The schematic diagram of electron–proton synergistic doping route. The whole system is immersed into an acid solution. The purple bubbles are free protons and the yellow column is a low-work function metal particle. **b** The mechanism of electron–proton synergistic doping effect: electrons flowing from metal with a higher Fermi level (i.e., lower work function $W_m$) to oxide with a lower Fermi level (i.e., higher work function $W_s$) at the interface. **c** The work function values for Zn and WO₃. **d** Experiments prove that Zn and WO₃ can realize the synergistic effect, which can be observed by eyesight due to the electrochromic effect. **e** Schematic depiction: this synergistic effect has a good spatial selection. **f** Sharp zinc pen was used to write "WO₃" in WO₃ film in acid solution. **g** This synergistic effect is reversible by 300 °C annealing in air.

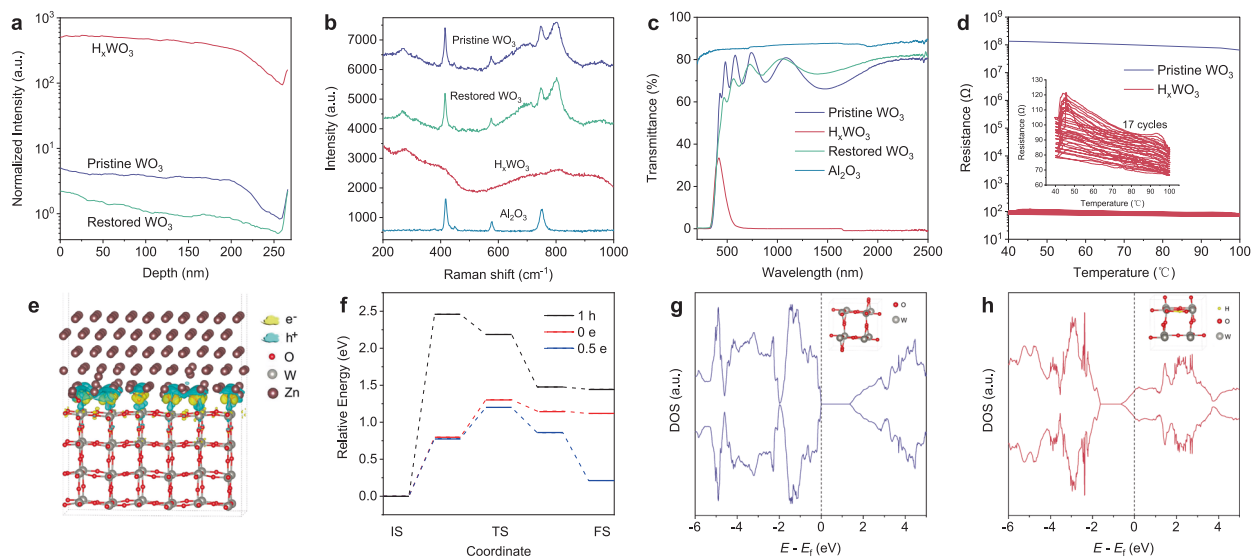

**Fig. 2 The characterizations and theoretical calculations for the electron doping WO₃ film. a** The secondary-ion mass spectrum (SIMS) tests for the WO₃ film after treated by electron–proton synergistic doping route, labeled by $H_xWO_3$. Then the $H_xWO_3$ annealed at 300 °C in air leads to the restored WO₃ sample. **b** The Raman peak at 802 cm⁻¹ for pristine WO₃ film is disappeared in $H_xWO_3$. While it can be restored after annealing the $H_xWO_3$ sample. **c** The UV–Vis-infrared transmission curves for the pristine WO₃, the $H_xWO_3$ and the restored WO₃ samples. **d** Resistance–temperature (R–T) results show the resistance of pristine WO₃ film is quite stable within the temperature from 40 to 100 degrees, while it will decrease with six orders of magnitude after the film was hydrogenated to $H_xWO_3$, indicating a pronounced metal–insulator transition (MIT). The inset is the R–T cycle test in air, showing the metallic $H_xWO_3$ is quite stable at ambient. **e** Computed differential charge distribution at Zn/WO₃ interface. Isosurface is 0.005 electrons per Å³. **f** Energy diagram along the reaction pathways of H migrating from WO₃ surface (initial state: IS) to subsurface (final state: FS) via the transition state (TS). Energy of IS is set as zero. **g, h** Density of states (DOS) of WO₃ (**g**) and H-doped WO₃ (**h**), together with inset photographs for the atomic models.

The mechanism of the quick insertion of H atom into $WO_3$ film by the zinc–acid treatment was also investigated by theoretical calculations. Considering the Fermi level difference for the $Zn/WO_3$ interface, electrons would flow from Zn to $WO_3$. The computed differential charge distributions of $Zn/WO_3$ interface confirmed the charge transfer behavior (Fig. 2e) as the model predicted in Fig. 1b, c. Bader charge analysis found ~0.07 electrons being transferred from each Zn atom to $WO_3$, leaving negative charges in the $WO_3$ part. The extra negative charges in $WO_3$ (donated from Zn metals) would attract the surrounding protons in acid solution to penetrate into the lattice. By simulating the reaction pathways of a hydrogen atom migrating from $WO_3$ surface to subsurface (Supplementary Fig. 8), we demonstrated that the H-migration barrier ($\Delta E_b$) is substantially lowered by increasing the number of negative charges (Fig. 2f). With charge conditions of 1 hole, neutral, and 0.5 electron, the $\Delta E_b$ value is 2.46, 1.32, and 1.20 eV, respectively. One can thus infer that electrons accumulated in the $WO_3$ can help proton diffusion into the lattice, and the encounter of protons and electrons inside the lattice completes the hydrogenation of $WO_3$, which was confirmed by the SIMS results in Fig. 2a. Thus, the same mechanism happened in $VO_2$ system is suitable for $WO_3$ system[17]. Importantly, H-doping induces significant changes to the electronic structures of $WO_3$. The calculated density of states in Fig. 2g, h show that the hydrogenation effective shifts up the $E_F$, closing the original bandgap in $WO_3$ and consequently bestowing metallic features to the doped system of $H_xWO_3$, which are responsible for the MIT (Fig. 2c) and low transmissivity of visible light and infrared (Fig. 2d) in our experiments. These calculated results are consistent with previous reports[11,16,29].

**Production of $WO_3$-based gratings.** Because the observed synergistic doping effect in $WO_3$ film was quite fast and the doping area was selective at ambient, some interesting patterns could be fabricated by combining with UV lithography. Fortunately, photoresist usually made by photopolymer materials are often stable in acid[30]. In addition, the pristine $WO_3$ film and $H_xWO_3$ film showed pronounced transmission variation in visible and near-infrared range (Fig. 2c), so optical gratings are possibly produced. Thus, at first, we prepared a macroscopic pattern on a $WO_3$ film as shape of "USTC" by a photoresist layer. It was found that the covered parts (the four characters) kept the original $WO_3$ phase, while the other areas were all hydrogenated, showing the fact that the photoresist layer can hinder the synergistic effect (Fig. 3a). A demonstration in Supplementary Video 4 recorded the diffusion around the covered photoresist. Furthermore, a set of microscopic bars of photoresist were fabricated by UV lithography in Fig. 3b. Then the bare microchannels of $WO_3$ film can be "dyed" freely by moving and touching a Zn probe in acid, for example, the two microchannels were selected to be colored (Fig. 3b). A demonstration is in Supplementary Video 5, where the colorations spread very fast along the selected channels.

Based on the above results, it was suggested that the controllable optical grating devices could be produced in this way. The specific procedures were described in Supplementary Fig. 9. A bar array and a square array of photoresist were fabricated by standard UV lithography on two $WO_3$ films (Fig. 3c, d), respectively. The same periods were 16 μm for the bar array in the horizontal direction and for the square array in both horizontal and perpendicular directions. Figure 3g, h showed surface morphology of these two arrays respectively by a three-dimensional profilometer. The thickness of the covered photoresist layer on each line or each square was >2 μm. Actually, these periodic arrays can be considered as typical relief gratings proved by diffraction measurement (setup of the probe device in Supplementary Fig. 10), the corresponding spots appearing in Fig. 3k, l, respectively.

While it treat these two arrays by the synergistic doping, the areas without coated photoresist layer were dyed as shown in Supplementary Fig. 11. After lift-off, the pattern of bars and squares were made in $WO_3$ films as shown in Fig. 3e, f, respectively (extended results in Supplementary Fig. 12). The contrast of the patterns should originate from the difference of metal and insulator areas. The diffraction spots patterns were recorded in Fig. 3m, n, and quite consistent with the results in Fig. 3k, l from the gratings covered photoresist layer. These results indicated our synergistic effect was an alternative technology for grating fabrication. In addition, we should emphasize that this kind of grating is different from the common relief gratings with physical grooves, such as shown in Fig. 3c, d, because these gratings are composed by the periodic metal and insulator $WO_3$ area with smooth surface (Fig. 3i, j). Because of this, we named it "planar grating". More importantly, this planar grating can be restored and reprocessed, since these periodic patterns could be removed by annealing in air. Indeed, the diffraction spots were disappeared as shown in Supplementary Fig. 13.

Traditional relief gratings usually have a fixed period. Here, we developed a kind of relief grating with reversibly tunable periods. By combining UV lithography and reactive ion etching (RIE), a $WO_3$ relief grating of 16 μm period was made in Fig. 3o, the surface morphology in Fig. 3p and the related diffraction spots (period: ~13.3 mm) in Fig. 3t. We immersed this grating in dilute $H_2SO_4$ solution. Then the original period could be easily doubled to be 32 μm by a Zn probe contacting the bars alternatively, since the touched $WO_3$ bars were hydrogenated and dyed quickly. Resultantly, the period of diffraction spots was half of the original period as shown in Fig. 3u. While if only a quarter of $WO_3$ bars were dyed (Fig. 3r), which means a new period of 64 μm, the period of diffraction spots was 1/4 of the original one (Fig. 3v). These results are consistent with the reciprocal law of grating diffraction[2]. To further verify the observation, another way to realize 64 μm period, dyeing three quarters of $WO_3$ bars, was also conducted in Fig. 3s. Indeed, the diffraction spots in Fig. 3w showed the same period as that in Fig. 3v. The extended photos of the colored gratings in different periods were exhibited in Supplementary Fig. 14, indicating the flexible period selection in $WO_3$ relief grating by the synergistic doping. More importantly, it should be emphasized again that all the processed gratings can be restored by annealing as shown in Supplementary Fig. 15, making this technique suitable for rewritable grating device fabrication in the future.

**Low scattering performance of the planar grating.** For typical relief gratings, scattering originates from random errors and irregularities of surfaces and sides, which cannot be avoided due to the physical grooves. In contrast, our planar grating with smooth surface can basically survive in these scattering. Figure 4a, b showed the derived diffraction intensity of relief grating coated photoresist and planar grating corresponding to Fig. 3l, n, respectively. It was observed that the background scattering in Fig. 4a was unambiguously stronger than that of Fig. 4b. The intensity along the dashed lines in Fig. 4a, b was plotted in Fig. 4c correspondingly, the red curve showing better quality not only lower scattering (blue dots), but also narrower width of peaks (inverted triangles). The calculated result indicated 21.3% scattering was suppressed in the planar grating compared with that of relief grating. Similar results also appeared in Fig. 4d, e, 11.3% suppression of scattering in Fig. 4f. In fact, the proposed planar grating can be considered as a kind of new volume phase grating without physical grooves. The low scattering is also reported in volume phase holography diffraction gratings (VPH diffraction gratings)[18,19]. Possibly serious scattering in doping gratings caused by inhomogeneity of doping is far away from our planar grating because the synergistic doping is homogenous, since it happens quickly and has outstanding diffusion. It should be noticed the low scattering in our planar gratings is intrinsic rather than accidental results caused by different technics. After preparation of the photoresist arrays, subsequent steps were conducted to fabricate the planar gratings (Supplementary Fig. 9). However, the deservedly stronger scattering originating from accumulative errors did not observed in experiments. Finally, thanks to the smooth surface, the planar gratings should have better accidental resistant than typical relief gratings, whose grooves are easily damaged by fingerprints, aerosols, moisture, or the slightest contact with any abrasive material.

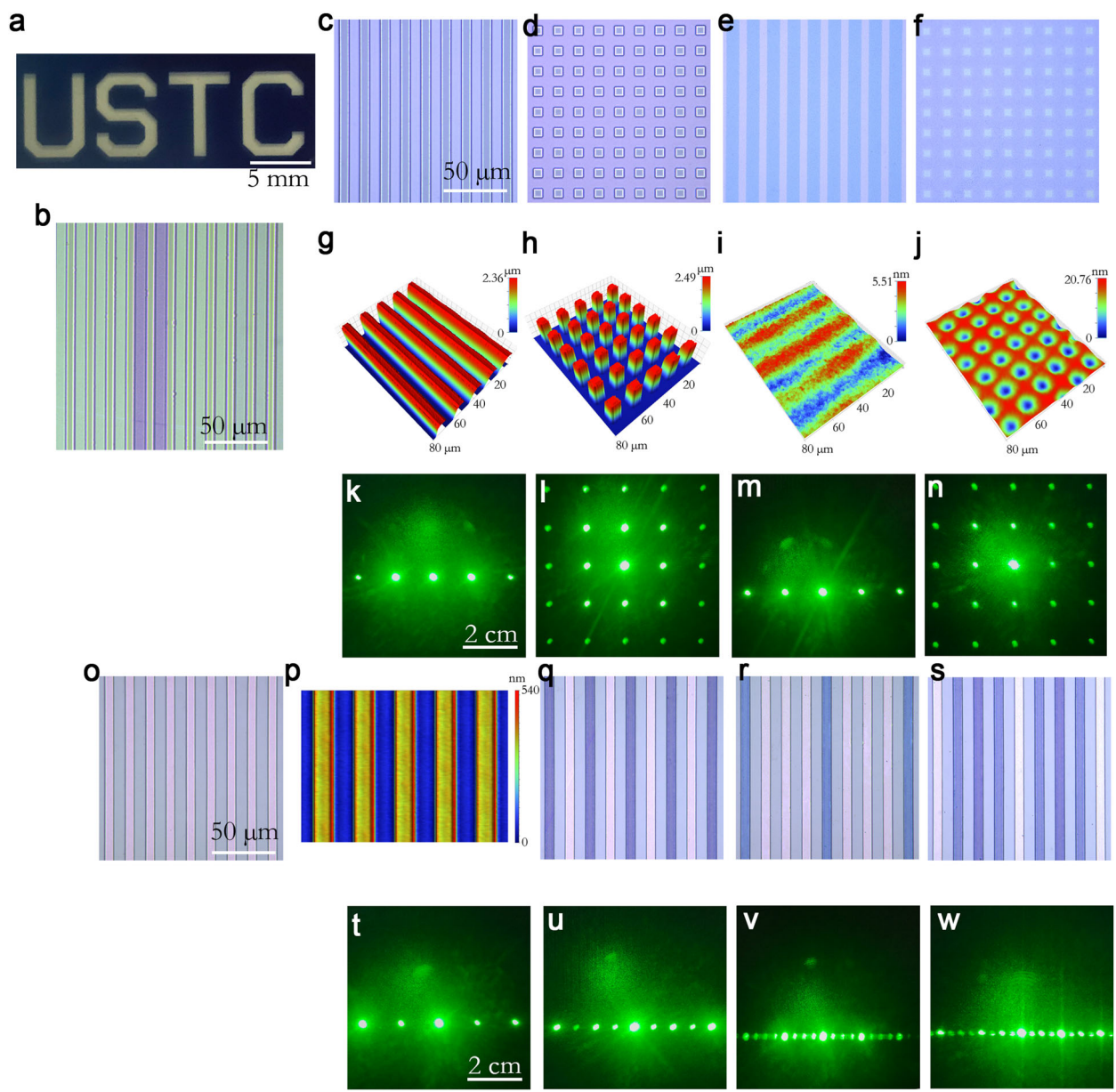

**Fig. 3 The performance of rewritable WO₃-based gratings. a** Photoresist is covered on a WO₃ film as the shape of "USTC", which proves that photoresist would hinder the electron–proton synergistic doping. **b** The synergistic effect is compatible with UV lithography, which can be conducted in microchannels by coating photoresist. **c–f** The photos recorded by optical microscope: a photoresist array in bar shape (**c**) and a photoresist array in square shape (**d**). After being processed by synergistic doping and lift-off, the corresponding pattern of bars (**e**) and squares (**f**) were obtained. Panels **c–f** have the same scale bar. The period is 16 μm. **g–j** The corresponding surface morphology of samples in panels **c–f**. **k–n** The corresponding diffraction spots of samples in panels **c–f**. For panels **k–n**, scale bar is the same. **o** The WO₃ relief grating in the period of 16 μm, **p** surface morphology, **t** diffraction spots. **q–s** Adjusting the period of the relief grating by the synergistic doping, 32 μm for panel **q**, 64 μm for panel **r**, 64 μm for panel **s**. **u–w** The corresponding diffraction spots of samples in panels **q–s**. For panels **o** and **q–s**, scale bar is the same. For panels **t–w**, scale bar is the same.

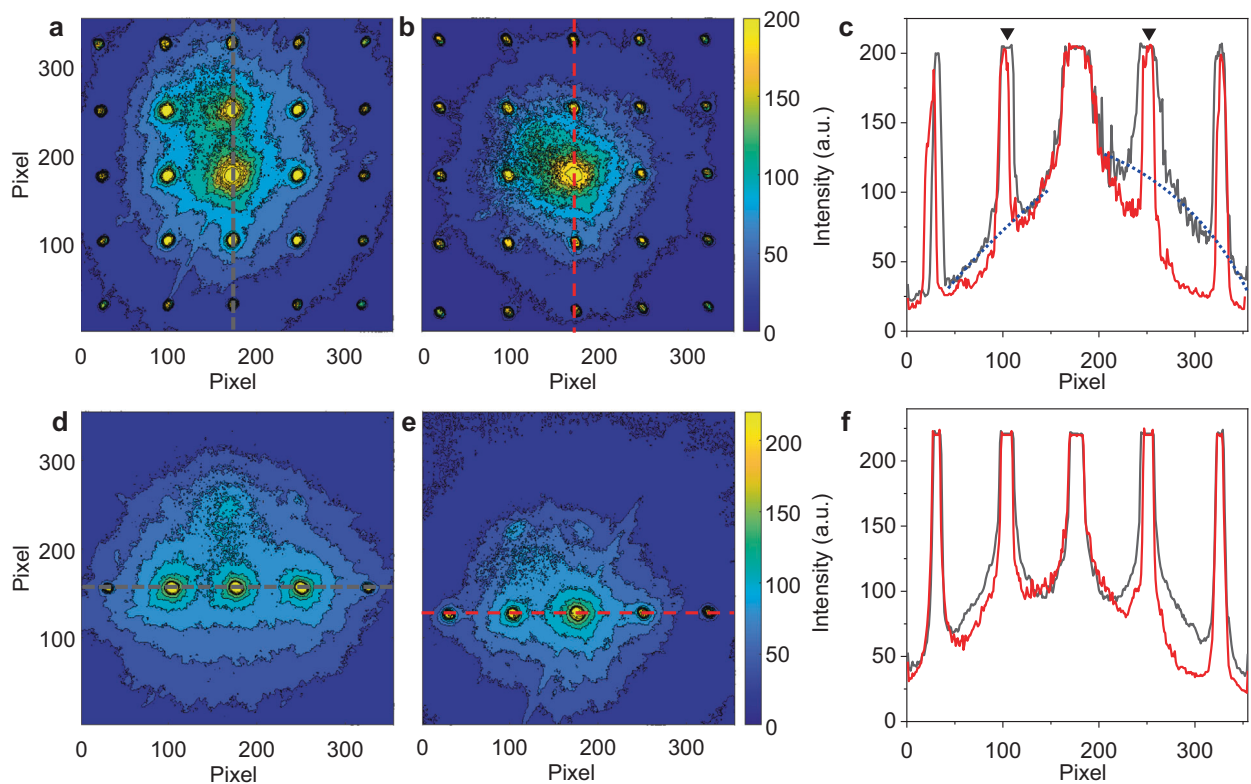

**Fig. 4 The analysis of scatter performance of the planar grating. a, b** Derived diffraction intensity of relief grating coated photoresist and planar grating from Fig. 3l, n, respectively. **c** Plotting the data along the dashed lines in **a** and **b** correspondingly. The red curve showing lower scattering (blue dots) and narrower width of peaks (inverted triangles) compared with the gray curve. **d-f** Same analysis applied on Fig. 3k, m.

## Conclusion

To conclude, we applied electron–proton synergistic doping to WO₃ film successfully and realized a direct visualization of insulator–metal transition in WO₃ film due to its pronounced electrochromic effect. This phase transition originated from the H atoms doping, which resulted in a stable metallic state of WO₃ film in air. We proved the synergistic effect was compatible with conventional ultraviolet photolithography. A kind of novel volume phase grating without physical grooves was developed, which showed better performance than typical relief gratings. The period of the WO₃ relief grating can also be tuned by selective hydrogenation. In order to evaluate the scattering performance, our experiment not only demonstrate a facile electron–proton synergistic doping strategy for oxide materials, but also supply a technique for tunable and rewritable grating fabrication in the future.

## Methods

**WO₃ film deposition.** WO₃ thin films were prepared on c-Al₂O₃ single-crystal substrates by reactive magnetron sputtering in an argon–oxygen atmosphere at 200 °C with a stoichiometric WO₃ target. Before loading the Al₂O₃ substrates into the chamber, they were cleaned with acetone and ethanol in an ultrasonic bath, and then rinsed several times with deionized water. The flow rates of argon and oxygen were fixed at 6.0 and 1.0 SCCM, respectively. The radio-frequency sputtering power during the deposition was maintained at 80 W. Under these conditions, the prepared WO₃ thin films were amorphous. The amorphous structure is in favor of the ions/protons insertion into WO₃ film and induced quick electrochromic results[31,32]. The final thickness of the prepared film was ~260 nm (Supplementary Fig. 6).

**Characterizations.** The purity of zinc particle and needle is >99.99%. The 2%wt H₂SO₄ solution was adopted in all experiments. The square resistance as a function of temperature was examined by an electric measurement system (ZJ2810B) with a variable temperature stage (Eurotherm 3504). The average rate of heating up is 0.15 K s⁻¹ and cool down by nature. Highly sensitive SIMS measurements (Quad PHI6600) were conducted to directly examine the hydrogen, oxygen, and aluminum concentration in each sample. H⁺ signals were normalized by the background signal in the substrate. The inductively coupled plasma optic emission spectrometer (iCAP 7400) was used to trace zinc concentration after the prepared HₓWO₃ was dissolved in hot H₂SO₄ solution. Encapsulating films in vacuum sealer bags by Deli 14886 (Vacuum Power: −85 kPa). To examine the crystal structure, XRD tests were carried out by Philips X'pert Pro, radiation source Cu Kα, λ = 0.15148 nm. The SEM images were recorded by XL-30 ESEM. Raman spectroscopy tests were recorded at room temperature by LabRAM HR Evolution. A 633 nm laser was used as the excitation source. The optical transmission was measured at room temperature by using a UV–Vis–IR spectroscopy (SolidSpec 3700). A home-made device was used for measuring the diffraction of the grating (Supplementary Fig. 10). The wavelength of the laser is 532 ± 10 nm. The max output power is <3000 mW. A three-dimensional profilometer (ContourGT-K 3D Optical Microscope) was used to measure surface morphology. An optical microscope (Leica DM8000) was used to record microimages. Photoresist arrays were made by standard UV photolithography. RIE was conducted by Oxford, Plasma Pro NGP 80. Hall measurement was carried on PPMS (DynaCool-14T). The Hall bar of HₓWO₃ was prepared by RIE through a silicon stencil. MATLAB R2019b was used to analyze the diffraction intensity. In order to evaluate the scattering performance, we calculated each integration (S) of the two curves in Fig. 4c (Fig. 4f), respectively, marked S₀ for gray curve, S₁ for red curve. The suppression of scattering (Δ) expressed as Δ = (S₁ − S₀)/S₀.

**Computational details.** All the calculations of the present work were performed by the Vienna Ab Initio Simulation Package with the density functional theory[33]. We chose the frozen-core all-electron projector augmented wave[34] model for core states and the Perdew−Burke−Ernzerh[35] for exchange and correlation functional. We set the kinetic energy cutoff of 400 eV for the plane-wave expansion of the electronic wave function. We set the force and energy convergence criterion to be 0.02 eV Å⁻¹ and 10⁻⁵ eV, respectively. The length of vacuum space was 20 Å to avoid interactions between periodic images. The solid–solid interfaces are built by attaching the Zn and WO₃ surface with mismatch <5%. The energy barrier was calculated as ΔE_b = E_TS − E_IS, where E_TS and E_IS represent the energies of the initial structures (IS) and the transition states (TS).

## Data availability

The data that support the findings of this study are available from the corresponding author upon reasonable request.

## References

1. Gaylord, T. K. & Moharam, M. G. Analysis and applications of optical diffraction by gratings. *Proc. IEEE* **73**, 894–937 (1985).
2. Palmer, C. A. & Loewen, E. G. *Diffraction Grating Handbook* (Newport Corporation, 2005).
3. Takeshima, N., Narita, Y., Tanaka, S., Kuroiwa, Y. & Hirao, K. Fabrication of high-efficiency diffraction gratings in glass. *Opt. Lett.* **30**, 352–354 (2005).
4. Rashid, I. et al. Wavelength-selective diffraction from silica thin-film gratings. *ACS Photonics* **4**, 2402–2409 (2017).
5. Zhong, L. et al. Fabrication of PCD micro cutting tool and experimental investigation on machining of copper grating. *Int. J. Adv. Manuf. Technol.* **88**, 2417–2427 (2016).
6. Lindau, S. The groove profile formation of holographic gratings. *Opt. Acta* **29**, 1371–1381 (2010).
7. Chen, S. et al. Gate-controlled VO2 phase transition for high-performance smart windows. *Sci. Adv.* **5**, eaav6815 (2019).
8. Shi, J., Zhou, Y. & Ramanathan, S. Colossal resistance switching and band gap modulation in a perovskite nickelate by electron doping. *Nat. Commun.* **5**, 4860 (2014).
9. Chen, X., Liu, L., Yu, P. Y. & Mao, S. S. Increasing solar absorption for photocatalysis with black hydrogenated titanium dioxide nanocrystals. *Science* **331**, 746–750 (2011).
10. Nishihaya, S. et al. Evolution of insulator-metal phase transitions in epitaxial tungsten oxide films during electrolyte-gating. *ACS Appl. Mater. Interfaces* **8**, 22330–22336 (2016).
11. Wang, M. et al. Electric-field-controlled phase transformation in WO3 thin films through hydrogen evolution. *Adv. Mater.* **29**, 1703628 (2017).
12. Leng, X. et al. Insulator to metal transition in WO3 induced by electrolyte gating. *npj Quant. Mater.* **2**, 1-7 (2017).
13. Xie, L. et al. Tunable hydrogen doping of metal oxide semiconductors with acid-metal treatment at ambient conditions. *J. Am. Chem. Soc.* **142**, 4136–4140 (2020).
14. Buch, V. R., Chawla, A. K. & Rawal, S. K. Review on electrochromic property for WO3 thin films using different deposition techniques. *Mater. Today* **3**, 1429–1437 (2016).
15. Yang, J. T. et al. Artificial synapses emulated by an electrolyte-gated tungsten-oxide transistor. *Adv. Mater.* **30**, e1801548 (2018).
16. Yao, X. et al. Protonic solid-state electrochemical synapse for physical neural networks. *Nat. Commun.* **11**, 3134 (2020).
17. Chen, Y. et al. Non-catalytic hydrogenation of VO2 in acid solution. *Nat. Commun.* **9**, 818 (2018).
18. Barden, S., Arns, J. A. & Colburn, W. S. in *Optical Astronomical Instrumentation, Pts 1 and 2*, Vol. 3355 (ed. Dodorico, S.) 866–876 (Spie-Int Soc Optical Engineering, 1998).
19. Barden Samuel, C., Arns James, A., Colburn Willis, S., Williams & Joel, B. Volume-phase holographic gratings and the efficiency of three simple volume-phase holographic gratings. *Publ. Astron. Soc. Pacific* **112**, 809–820 (2000).
20. Meyer, J. et al. Highly efficient simplified organic light emitting diodes. *Appl. Phys. Lett.* **90**, 113506 (2007).
21. Kröger, M. et al. Role of the deep-lying electronic states of MoO3 in the enhancement of hole-injection in organic thin films. *Appl. Phys. Lett.* **95**, 123301 (2009).
22. Meyer, J. et al. Transition metal oxides for organic electronics: energetics, device physics and applications. *Adv. Mater.* **24**, 5408–5427 (2012).
23. Altendorf, S. G. et al. Facet-independent electric-field-induced volume metallization of tungsten trioxide films. *Adv. Mater.* **28**, 5284–5292 (2016).
24. Crandall, R. S. & Faughnan, B. W. Electronic transport in amorphousHxWO3. *Phys. Rev. Lett.* **39**, 232–235 (1977).
25. Ramadoss, K. et al. Sign reversal of magnetoresistance in a perovskite nickelate by electron doping. *Phys. Rev. B*, **94**, 235124 (2016).
26. Chen, S. et al. Sequential insulator-metal-insulator phase transitions of VO2 triggered by hydrogen doping. *Phys. Rev. B* **96**, 125130 (2017).
27. ViolBarbosa, C. et al. Transparent conducting oxide induced by liquid electrolyte gating. *Proc. Natl Acad. Sci. USA* **113**, 11148–11151 (2016).
28. Jackson, J. D. *Classical Electrodynamics* (Wiley, 2007).
29. Hjelm, A., Granqvist, C. G. & Wills, J. M. Electronic structure and optical properties of WO3, LiWO3, NaWO3, and HWO3. *Phys. Rev. B* **54**, 2436–2445 (1996).
30. Crivello, J. V. & Reichmanis, E. Photopolymer materials and processes for advanced technologies. *Chem. Mater.* **26**, 533–548 (2013).
31. Kamal, H., Akl, A. A. & Abdel-Hady, K. Influence of proton insertion on the conductivity, structural and optical properties of amorphous and crystalline electrochromic WO3 films. *Phys. B* **349**, 192–205 (2004).
32. Taylor, D. J., Cronin, J. P., Allard, L. F. & Birnie, D. P. Microstructure of laser-fired, sol–gel-derived tungsten oxide films. *Chem. Mater.* **8**, 1396–1401 (1996).
33. Kresse, G. & Furthmüller, J. Efficiency of ab-initio total energy calculations for metals and semiconductors using a plane-wave basis set. *Comput. Mater. Sci.* **6**, 15–50 (1996).
34. Blöchl, P. E. Projector augmented-wave method. *Phys. Rev. B* **50**, 17953–17979 (1994).
35. Perdew, J. P., Burke, K. & Ernzerhof, M. Generalized gradient approximation made simple. *Phys. Rev. Lett.* **77**, 3865–3868 (1996).

## Acknowledgements

This work was partially supported by the National Key Research and Development Program of China (2016YFA0401004), users with Excellence Program of Hefei Science Center CAS, the National Natural Science Foundation of China (12074356), the funding supported by the Youth Innovation Promotion Association CAS, the Fundamental Research Funds for the Central Universities. This work was partially carried out at the USTC Center for Micro and Nanoscale Research and Fabrication. The authors also acknowledged the supports from the Anhui Laboratory of Advanced Photon Science and Technology. The approved beamtime on the XMCD beamline (BL12B) in National Synchrotron Radiation Laboratory (NSRL) of Hefei was also appreciated.

## Author contributions

Y.C. and C.Z. conceived the project. Y.C. fabricated the devices, performed the measurements, and analyzed the data. C.H., X.Z., and B.L. grew the WO3 films. C.H. performed the SEM, XRD, and Raman characterizations. H.R. performed the SIMS experiments. G.Z. contributed many valuable comments and ideas in revised manuscript. Y.C. and L.L. analyzed scattering performance of the planar grating. L.X and J.J. conducted the theoretical calculations. Y.C., J.J., and C.Z. wrote the paper, and all authors commented on it.

## Competing interests

The authors declare no competing interests.

## Additional information

**Supplementary information** The online version contains supplementary material available at https://doi.org/10.1038/s43246-021-00141-2.

**Correspondence** and requests for materials should be addressed to J.J. or C.Z.

**Peer review information** Primary handling editor: John Plummer

**Reprints and permission information** is available at http://www.nature.com/reprints

**Publisher's note** Springer Nature remains neutral with regard to jurisdictional claims in published maps and institutional affiliations.